\documentclass[a4paper,10pt]{article}

\usepackage{hyperref}
\hypersetup{
pdfpagemode=UseOutlines,      % UseOutlines, UseThumbs, None, FullScreen : agencement au démarrage
pdfstartview=Fit,             % Fit, FitH, FitB, FitBH : vue de la page au départ (pleine largeur...)
pdffitwindow=true,            % bool: Maximiser
pdfpagelayout=OneColumn,% SinglePage, TwoColumnsRight/Left, OneColumn : affichage des pages
pdftoolbar=true,              % bool: Affichage de la barre d'outils
pdfmenubar=true,              % bool: Affichage de la barre de menus
bookmarksopen=false,          % bool: Dépliement des signets
bookmarksnumbered=true,       % bool: Numérotation des signets
colorlinks=true,              % bool: Liens colorés
pdfauthor={Florent Michel, Renaud Parentani},     % Auteur
pdftitle={},    % Titre
pdfcreator=PDFLaTeX,          % 
pdfproducer=PDFLaTeX,         %
linkcolor=blue,               % Couleur des liens
urlcolor=blue,                %             url
anchorcolor=black,            %         du texte 
citecolor=blue,              % Couleur de citation 
frenchlinks=false,             % bool: Utiliser des petites majuscules pour les liens, plutôt que de la couleur
pdfborder={0 0 0}             % Ne pas encadrer les liens
}

\usepackage{color}
\usepackage{graphicx}
\usepackage{amsmath,amssymb,amsfonts,amsthm}
\usepackage{slashed}
\usepackage[utf8]{inputenc}   
\usepackage{amssymb}
\usepackage{mathrsfs}
\usepackage{multido}
\usepackage[english]{babel}
\usepackage[affil-it]{authblk}
\usepackage{float}

\newcommand{\be}{\begin{eqnarray}} 
\newcommand{\ee}{\end{eqnarray}}
\newcommand{\om}{\ensuremath{\omega}}

\newcommand{\pd}{\ensuremath{\partial}}

\newcommand{\lp}{\ensuremath{\left(}}
\newcommand{\rp}{\ensuremath{\right)}}

\newcommand{\Fmax}{\ensuremath{F_{\rm max}}}
\newcommand{\ommax}{\ensuremath{\om_{\rm max}}}
\newcommand{\ommin}{\ensuremath{\om_{\rm min}}}
\newcommand{\hmin}{\ensuremath{h_{\rm min}}}
\newcommand{\hmax}{\ensuremath{h_{\rm max}}}
\newcommand{\cmin}{\ensuremath{c_{\rm min}}}
\newcommand{\eq}[1]{Eq.~(\ref{#1})}

\newcommand{\Fig}[1]{Fig.~\ref{#1}}

\usepackage[titletoc,title]{appendix} %F Pour afficher "Appendix" devant les titres des annexes
\newcommand{\writeadot}{\hspace*{-0.5 cm}:\hspace*{0.4 cm}} %Pour ajouter deux points dans le titre des appendices

\begin{document}

\title{Mode mixing in sub- and trans-critical flows over an obstacle: \\ 
When should Hawking’s predictions be recovered?} 
\author{Florent Michel and Renaud Parentani}
%\address{Laboratoire de Physique Th\'eorique, CNRS UMR 8627, B\^atiment 210, Universit\'e Paris-Sud 11, 91405 Orsay Cedex, France} %F à décommenter pour la version WS
\affil{\small Laboratoire de Physique Th\'eorique, CNRS UMR 8627, B\^atiment 210, \\
Universit\'e Paris-Sud 11, 91405 Orsay Cedex, France}

%\bodymatter %F à échanger pour la version WS
\date{} %F À décommenter dans la version arXiv pour ne as afficher la date (et éviter un possible conflit avec celle donnée par arXiv)
\maketitle 

\begin{abstract}
We reexamine the scattering coefficients of shallow water waves blocked by a stationary counter current over an obstacle. By considering series of background flows, we show that the most relevant parameter is $F_{\rm max}$, the maximal value of the ratio of the flow velocity over the speed of low frequency waves. For subcritical flows, i.e., $F_{\rm max} < 1$, there is no analogue Killing horizon and the mode amplification is strongly suppressed. Instead, when $F_{\rm max} \gtrsim 1.1$, the amplification is enhanced at low frequency and the spectrum closely follows Hawking's prediction. We further study subcritical flows close to that used in the Vancouver experiment. Our numerical analysis suggests that their observation of the  ``thermal nature of the mode conversion'' is due to the relatively steep slope on the upstream side and the narrowness of the obstacle. 
\end{abstract}

\medskip
\medskip
\medskip

\section{Introduction} 

In~\cite{Unruh:1980cg,Unruh:1994je}, W.~Unruh pointed out that one could observe the Hawking emission  using analogue systems mimicking the scattering of light on a black hole metric. His argument rests on the possibility to observe the scattering of linear waves propagating against a transcritical counter flow. Since then several theoretical and experimental works have made this idea more concrete. Let us present here the most relevant results for the scattering of shallow water waves~\cite{Schutzhold:2002rf}.  

The first question that was addressed concerns the spectral properties of the analogue Hawking radiation taking short distance dispersion into account. It was found~\cite{Unruh:1994je,Brout:1995wp,Corley:1996ar} that the spectrum is robust. That is, the relative deviations with respect to the Planck spectrum at the Hawking temperature $\kappa/2\pi$ are linear in $\kappa/\Lambda$, where $\kappa$ is the surface gravity of the hole, and $\Lambda \gg \kappa$ is the dispersive high momentum scale. In these works, the stationary background flow was {\it assumed} to be transcritical. This guarantees that long wave-length waves effectively propagate in an analogue space-time metric which possesses a Killing horizon.~\footnote{Subsequent works~\cite{Macher:2009tw,Finazzi:2012iu,Coutant:2011in,Robertson:2012ku} showed that the leading deviations scale as $(\kappa/\Lambda)\times D^{-3/2}$ (for quartic dispersion), where $D = \kappa x_{\rm lin}/c_{\rm H}$ characterizes the spatial extension $x_{\rm lin}$ of the near horizon region where the gradients of the flow can be treated as constants. $c_{\rm H}$ is the velocity of the low frequency waves evaluated on the horizon. Hence, to avoid large deviations one should work with $D \gtrsim 0.1$, i.e., the flow must be clearly subcritical on one side and supercritical on the other side of the horizon.}
 
The second question concerns the actual properties of the background flows that have been realized in experiments aiming at detecting the analogue Hawking effect. It appears that the background flows in water waves experiments~\cite{Rousseaux:2007is,Weinfurtner:2010nu} (and probably also those using light in non-linear media~\cite{Belgiorno:2010wn,Schutzhold:2010am}) were {\it not} trans-critical. To understand the observations, one must thus start the analysis afresh and theoretically calculate the spectral properties in sub-critical flows. 

Following the procedure of~\cite{Michel:2014zsa}, we study the scattering coefficients in a series of flows where $F_{\rm max}$, the maximal value of the ratio of the flow velocity over the speed of low frequency waves, 
ranges from $1.25$ (transcritical) down to $0.75$ (subcritical). Our first aim is to display the key role played by  $F_{\rm max}$ in dividing the behavior of scattering coefficients between trans- and sub-critical flows. For trans-critical flows with $\Fmax \gtrsim 1.1$, we recover Hawking's prediction up to small corrections. The scattering on subcritical flows instead does not follow simple laws. Keeping $F_{\rm max} \simeq 0.7$ fixed, our second aim is to identify the {\it subdominant} role played by the length and the two slopes of the obstacle. For such flows, we show that changing any of these three parameters significantly affects the scattering coefficients. Finally, we compare our results to those obtained from the flow used in the Vancouver experiment~\cite{Weinfurtner:2010nu}. Our findings suggest that the reported observation of the ``thermal nature of the mode conversion'' is related to the specific properties of their obstacle. 

\section{Scattering coefficients in sub- and trans-critical flows}

We present the essential steps to compute the scattering coefficients of shallow water waves sent against a stationary flow.~\footnote{In this work, we only consider flows without undulation, that is, without the zero-frequency modulation of the free surface which is generally present in the downstream region. In Section III.B.3 of~\cite{Michel:2014zsa}, we numerically found that the relative modifications of the scattering coefficients are of the order of $\delta h /\hmin$, where $\hmin$ is the minimum water height found on top of the obstacle, and $\delta h$ the amplitude of the undulation. For the flow of~\cite{Weinfurtner:2010nu}, we estimate that $\delta h /\hmin \sim 0.1$. Hence, the undulation should not play any significant role for this type of flows. The reader interested in the spectral modifications induced by undulations can consult Ref.~\cite{Busch:2014hla}. Although that work deals with flowing atomic Bose condensates, we checked that its main results also apply to water waves. This was expected in virtue of the similarities of the Gross-Pitaevski and Korteweg-de Vries equations. \label{undul}} More detailed explanations can be found in~\cite{Michel:2014zsa}. 

When the fluid is ideal, inviscid, and incompressible, the propagation of surface waves on an inhomogeneous, laminar, two-dimensional flow is (approximatively) governed by~\cite{Schutzhold:2002rf,Michel:2014zsa,Unruh:2012ve,Coutant:2012mf} 
\be \label{eq:waveeq}
\left[ \left(\partial _t+\partial _x v(x)\right)\left(\partial _t+v(x)\partial _x\right) -i g \partial _x \tanh \left(-i h(x) \partial _x\right)\right] \phi(t,x) = 0, 
\ee 
where the partial derivatives act on all factors on their right. Here $v(x)$ is the horizontal flow velocity, $h(x)$ the background fluid depth, and $g$ the gravitational acceleration. The field $\phi(t,x)$ is related to the linear variation of the free surface by $\delta h(t,x) = -\frac{1}{g}\left(\partial _t+ v \partial _x\right)\phi(t,x)$. $\delta h$ is the quantity measured in the experiments~\cite{Rousseaux:2007is,Weinfurtner:2010nu,Euve:2014aga}.

Since the background flow is stationary, we work at fixed frequency $\om$, with stationary modes $e^{- i\omega t}\phi_\omega(x)$. As in~\cite{Michel:2014zsa}, we work in a {\it weak dispersive regime} where \eq{eq:waveeq} becomes 
\be \label{eq:om}
\left[ 
\left(-i \omega +\partial _xv\right)\left(-i \omega +v\partial_x\right) 
- g \partial_x h \partial_x 
-\frac{g}{3}\partial_x 
\left(h \partial_x \right)^3 
\right]
\phi_\omega(x) 
= 0 .
\ee
In the long wave length approximation, the quartic term is also neglected. Up to a conformal factor, the resulting equation has the form of a two dimensional d'Alembert equation in a curved metric given by $ds^2 = -c^2 dt^2 + (dx - v dt)^2$, where $c(x)^2 = g h(x)$ is the speed of low-frequency waves. The metric possesses a Killing horizon if $c(x)= |v(x)|$ at some point, i.e., if the Froude number $F \doteq |v/c|$ crosses 1. Its surface gravity is given by $\kappa = c \left\lvert \partial_x F \right\rvert$ evaluated where $F = 1$. 

In stationary flows, $v(x) = J/h(x)$, where $J$ is the conserved current. Hence the stream is completely characterized by $h(x)$ and $J$. In what follows we study a series of flows parameterized by their minimum water depth $h_{\rm min}$ reached on top of the obstacle. When $h_{\rm min} < h_c = (J^2/g)^{1/3}$, $F_{\rm max} = (h_c/h_{\rm min})^{3/2}$ is larger than $1$ and the flow is transcritical. Such a flow possesses a black hole horizon on the upstream slope and a white hole horizon on the downstream one. Instead, when $F_{\rm max}< 1$, the flow remains globally subcritical. Importantly, \eq{eq:om} offers a reliable approximation when the non-homogeneities of the flow are essentially localized in a region where $F \approx 1$. Hence \eq{eq:om} can be used to study the changes of the scattering coefficients when $F_{\rm max}$ decreases from $\sim 1.2$ down to $\sim 0.7$. 

\begin{figure}[h]
\begin{center}
\includegraphics[width = 0.6 \linewidth]{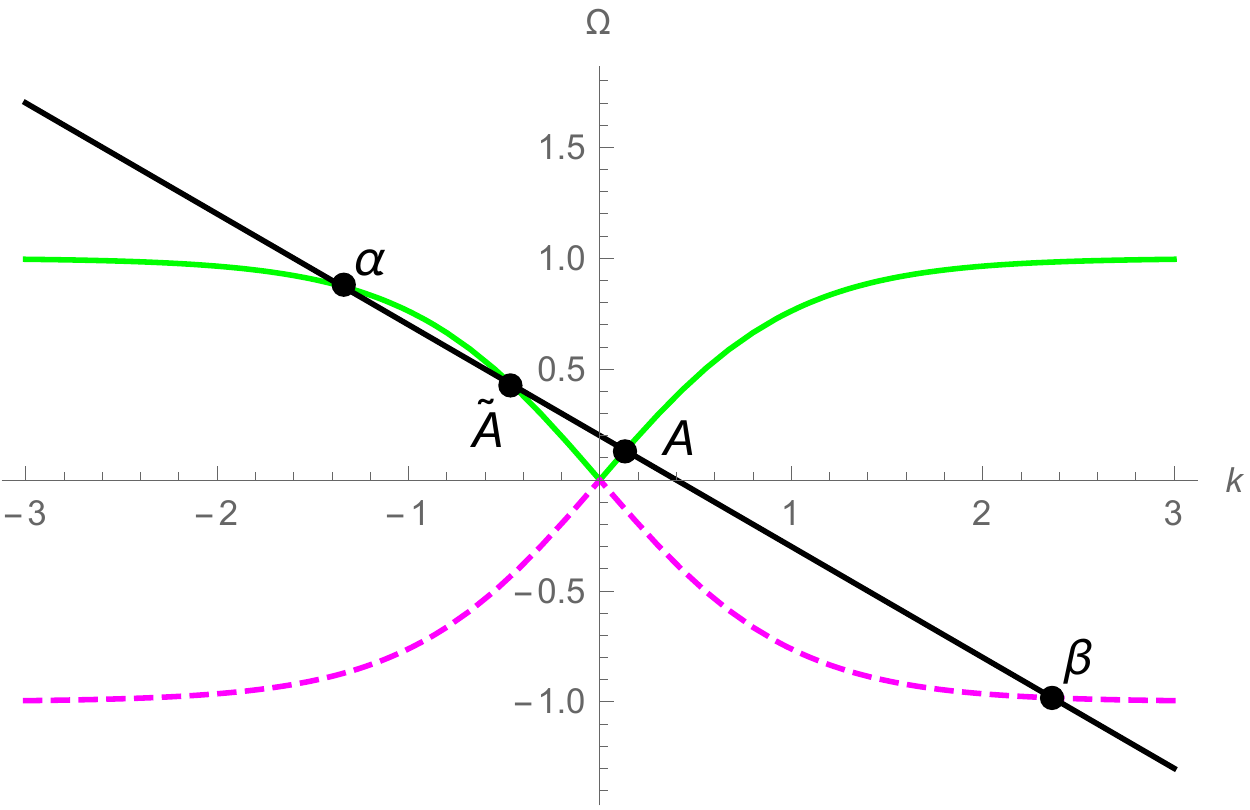} 
\end{center}
\caption{We show the dispersion relation associated with \eq{eq:waveeq} in the fluid frame, $\Omega^2 = g k \tanh(hk)$. The green continuous (purple dashed) curves describe the branches with positive (negative) values of $\Omega$. The straight line gives $\Omega = \om - vk$ evaluated in a subcritical flow to the right $0 < v < c = \sqrt{g h}$ 
and for a fixed $\om > 0$. The black dots indicate the four real roots $k_\om$. They are labelled by the coefficient of the corresponding outgoing wave of \eq{eq:Bsub}. Notice that only the last root lives on the negative $\Omega$ branch.}
\label{fig1}
\end{figure}

Scattering of incident counterpropagating waves $\phi_\omega^{\leftarrow, in}$ was observed in~\cite{Rousseaux:2007is,Weinfurtner:2010nu,Euve:2014aga}. For $v > 0$, these waves propagate to the left, as indicated by the exponentiated arrow. When scattered, they  give rise to four outgoing waves: 
\be 
\label{eq:Bsub}
\phi_\omega^{\leftarrow, in} \to \alpha_\omega \, \phi_\om^{\rightarrow,d,out} + \tilde{A}_\omega \, \phi_\omega^{\leftarrow, out} + A_\omega \, \phi_\om^{\rightarrow, out} + \beta_\omega \lp \phi_{-\om}^{\rightarrow,d,out} \rp^*.
\ee
Each outgoing wave is unambiguously identified by its wave-vector in an asymptotic region, see Fig.~\ref{fig1}. 

Starting from the left, the first root describes $\phi_\om^{\rightarrow,d,out}$, a dispersive short wave number mode propagating to the right. (By dispersive, we mean that this root exists only because the dispersion relation is not linear.) The second root describes a low wave number mode propagating to the left. When evaluated on the left (upstream) side, it describes the transmitted wave $\phi_\om^{\leftarrow, out}$. On the right side it describes the incident wave $\phi_\omega^{\leftarrow, in}$. The third root describes the elastically scattered mode which also propagates to the right. As we shall see, this long wave length mode plays no significant role in the scattering. The fourth root instead is essential. It describes the other dispersive short wave number mode propagating to the right. Unlike the other modes, it carries a negative energy and has a negative norm.\footnote{The conserved scalar product of two solutions of \eq{eq:waveeq} is given by~\cite{Unruh:1994je}
\be \label{eq:Kprod}
\lp \phi_1, \phi_2 \rp \equiv i \int \lp \phi_1^* (\pd_t + v \pd_x) \phi_2 - \phi_2 (\pd_t + v \pd_x) \phi_1^* \rp \, dx .
\ee 
Since $\lp \phi_1 , \phi_1 \rp$ is not positive definite, the scattering can lead to a mode amplification while preserving the norm. This mechanism is also known as over-reflection~\cite{OnOverReflection}, and is at the root of the (analogue) Hawking effect.} It has been complex conjugated so that all modes $\phi_{\pm \om}^{in/out}$ have a positive unit norm.

Using these normalized modes, the scattering coefficients obey 
\be \label{eq:4coef}
\left\lvert \alpha_\omega \right\rvert^2 + | \tilde{A}_\omega |^2 + \left\lvert A_\omega \right\rvert^2 - \left\lvert \beta_\omega \right\rvert^2 = 1. 
\ee
Notice that one recovers the standard relation  $\left\lvert \alpha_\omega \right\rvert^2 - \left\lvert \beta_\omega \right\rvert^2 = 1 $ iff $|\tilde{A}_\omega |^2 + |A_\omega |^2 \ll 1$. In relativistic settings, when neglecting the gray body factor (the equivalent of $|A_\omega|^2$), it is fulfilled and the radiation emitted from a Killing horizon follows a Planck law at the Hawking temperature: $\left\lvert \beta_\omega \right\rvert^2 = (e^{2 \pi \om/\kappa} - 1)^{-1}$. Importantly, when $\Fmax > 1$, this result is recovered from \eq{eq:om} in the dispersionless limit $h\to 0$ at {\it fixed} $v(x)$ and $c(x)$, where $\kappa$ is given by the expression below \eq{eq:om}. As shown in~\cite{Macher:2009tw}, the leading deviations are proportional to $h$ which here gives the dispersive short distance scale.

\subsection{Dependence on $\Fmax$}

\begin{figure}[h]
\begin{center}
\includegraphics[scale=0.6]{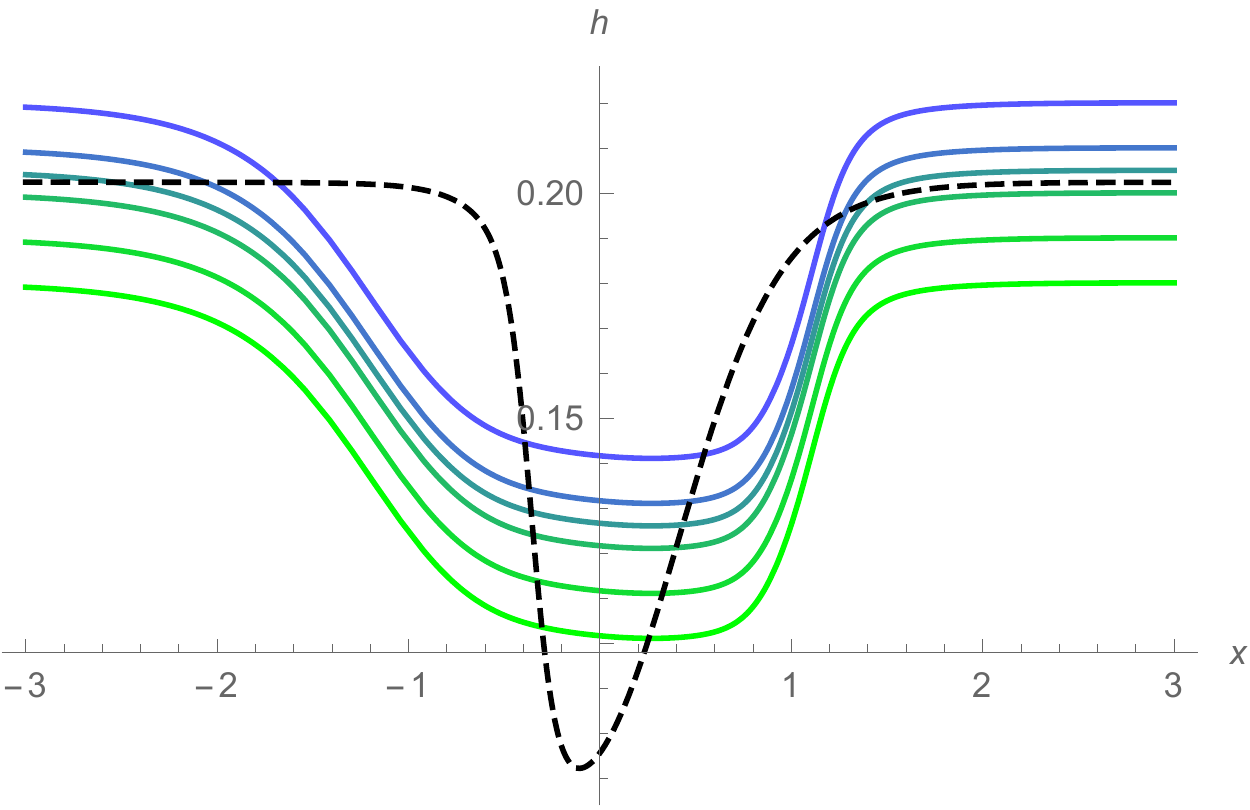} $\quad$ 
\includegraphics[scale=0.6]{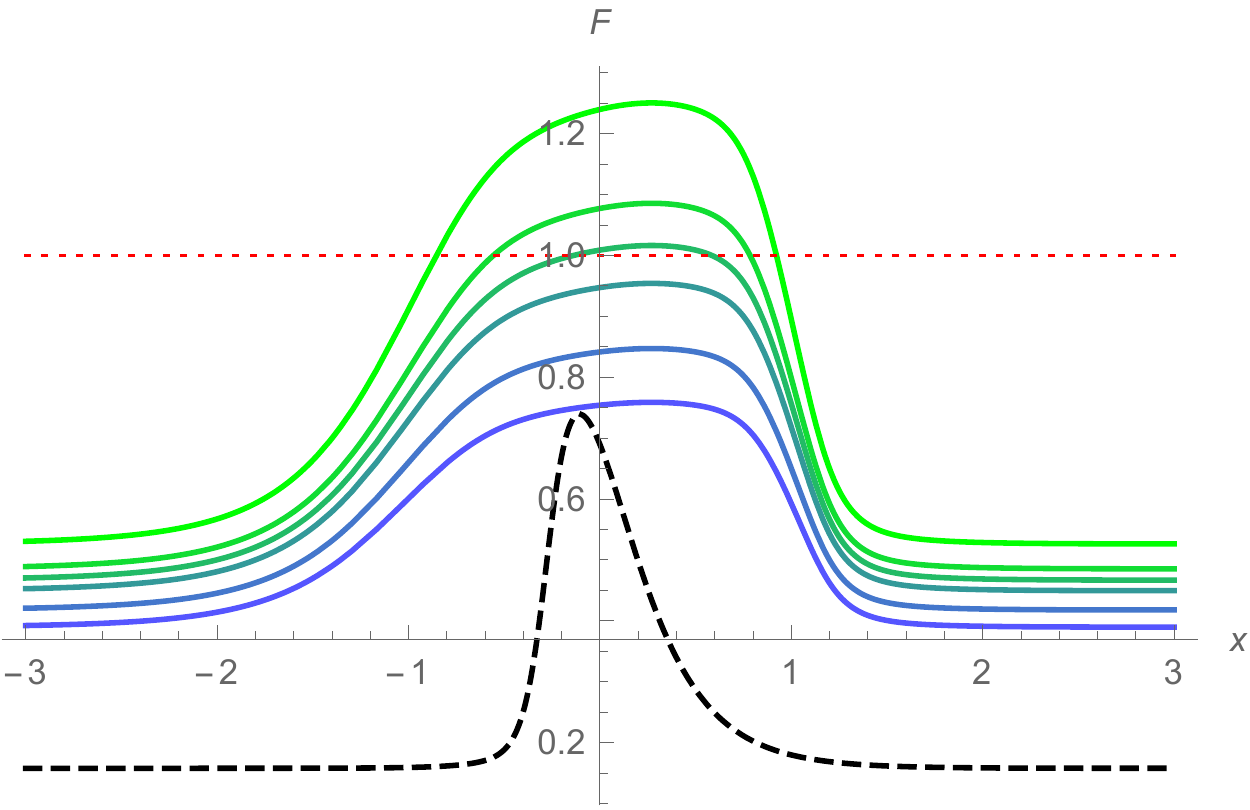}
\end{center}
\caption{We show the water height $h(x)$ (left) and the corresponding Froude number $F(x) = v(x) / c(x)$ 
(right) for 6 different flows to the right. The unit of the horizontal (and vertical for $h$) axis is the meter. The narrower profiles in dashed lines give our estimations of $h$ and $F$ for the flow used in the Vancouver experiment~\cite{Weinfurtner:2010nu}. The horizontal dotted line on the right panel gives the critical value $F = 1$. We see that three upper flows are transcritical, whereas the other ones are subcritical. Notice that the undulation for the flow of~\cite{Weinfurtner:2010nu} is not shown, see footnote~\ref{undul}.} 
\label{fig2}
\end{figure}

We now study the behavior of the four scattering coefficients when ``removing'' the Killing horizon by lowering $F_{\rm max}$ below 1. In  Fig.~\ref{fig2} we show the series of 6 flows we shall use. These flows interpolate from $F_{\rm max} = 1.25$ down to $0.75$. The profiles of $h(x)$ and $F(x) = (h_c/h(x))^{3/2}$ are given for a fixed value of the current $J = 0.126 \, {\rm m^2/s}$. The intermediate flow with $\Fmax \simeq 0.95$ has been obtained by solving nonlinear hydrodynamical equations, following the method presented in App. A of~\cite{Michel:2014zsa}. The 5 other flows have been obtained from it by adding a constant value to the water depth. It should be noticed that their downstream slope (on the right) is steeper than their upstream one. 

It should also be noticed that these flows radically differ from that of Ref.~\cite{Weinfurtner:2010nu}. As can be seen in Fig.~\ref{fig2}, the latter is much more narrow and has an higher slope in its upstream side, while it has typically the same slope on the downstream side. Its characteristic length scales are $\hmin \sim 8 {\rm cm}$ and $\hmax \sim 20 {\rm cm}$, and the corresponding current is $J = 0.045 {\rm m^2/s}$. As we shall see in the second part of our analysis, its narrower character does affect the spectra, although less importantly than the value of $\Fmax$.~\footnote{Given the available information and the complexity of hydrodynamical flows, it is difficult to give a precise estimate of $\Fmax$ for the experiment of Ref.~\cite{Weinfurtner:2010nu}. The Bernouilli equation in the low-gradient approximation gives $\Fmax \approx 0.5$. Taking into account the dispersive term from the Korteweg-de Vries equation gives $\Fmax \approx 0.65$, close to the value used in~\cite{Michel:2014zsa}. Yet this value is rather sensitive to the parameterization of the obstacle one adopts. In brief, we estimate that $0.6 \lesssim \Fmax \lesssim 0.75$. The validity of this range is strongly supported by the good agreement with the observations of the transmission coefficient $\tilde A_\om$ performed in~\cite{Euve:2014aga}. In Fig.~\ref{fig2} and forthcoming simulations, we work with $\Fmax \approx 0.75$.}

\begin{figure}[h]
\begin{center}
\includegraphics[scale=0.50]{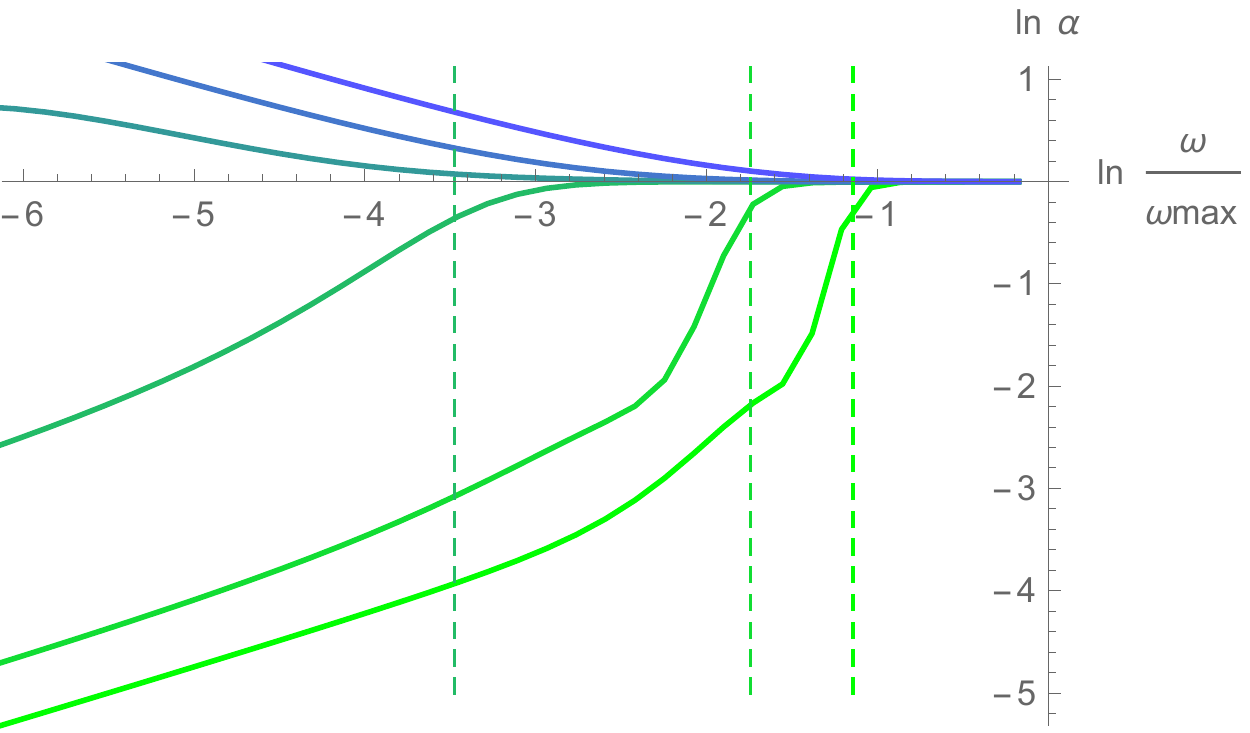} 
\includegraphics[scale=0.50]{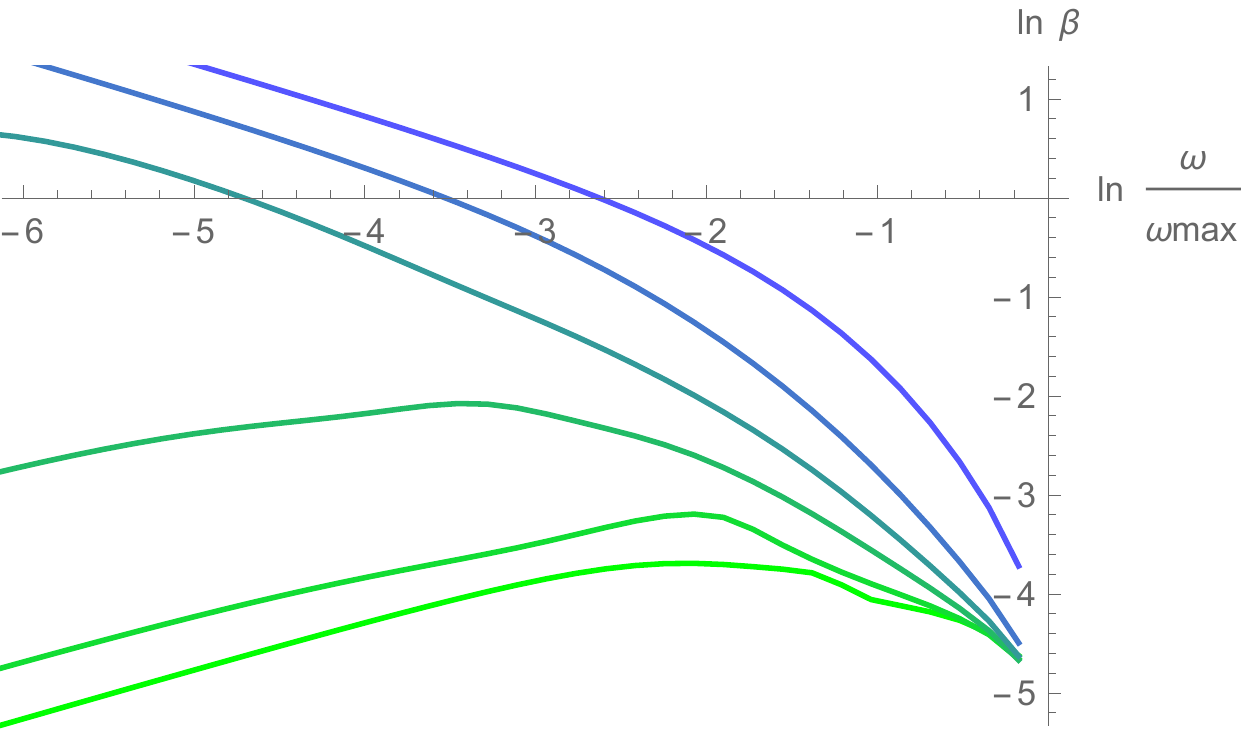}
\medskip
\includegraphics[scale=0.50]{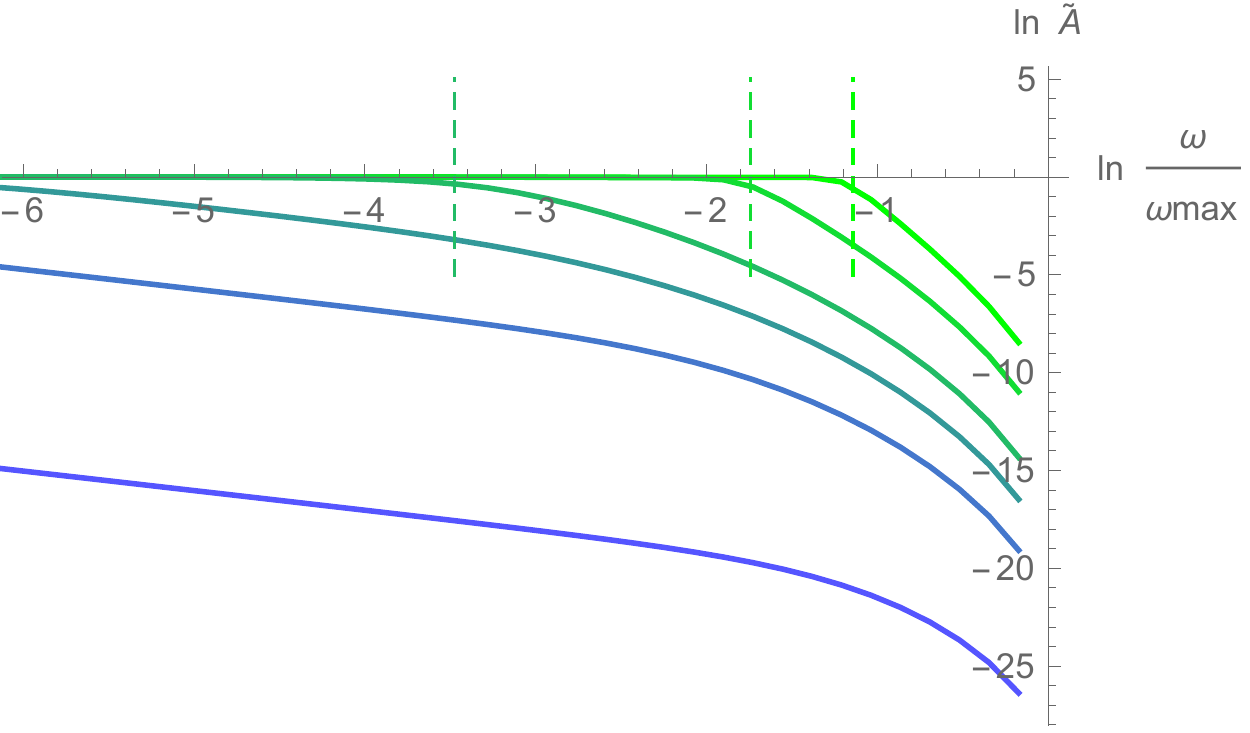}
\includegraphics[scale=0.50]{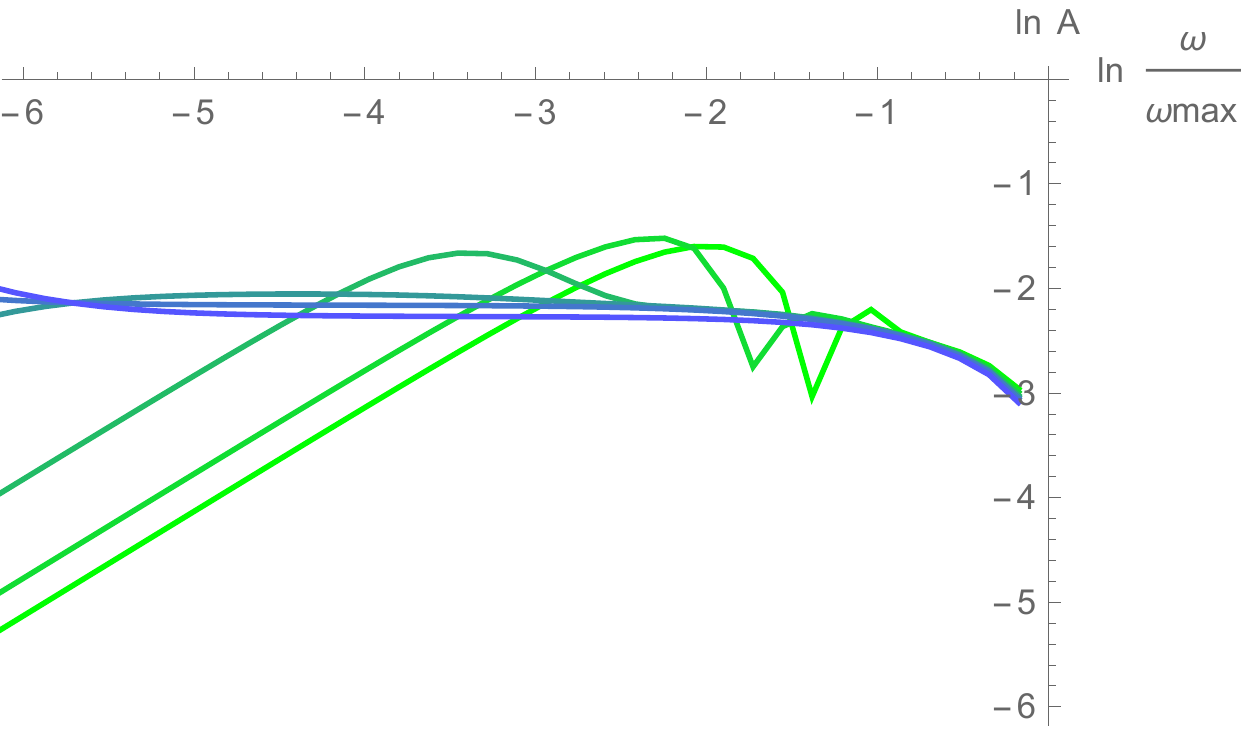}
\end{center}
\caption{The logarithms of the four scattering coefficients associated with the above series of flows are represented as functions of $\ln \frac{\om}{\ommax}$. From the two upper plots one sees that the behavior of $|\alpha_\om|$ (left) and $|\beta_\om|$ (right) principally depends on the trans- (the three upper curves) or the sub-critical character of the flow. In the left lower plot, for the three subcritical flows, one sees that the norm of the transmission coefficient approaches $1$ (for decreasing values of $\om$) near the critical frequency where the norm of $\alpha_\om$ starts diminishing. The three vertical dashed lines indicate the values of $\ommin$, see \eq{ommin}. In the right lower plot, we see that the norm of the reflexion coefficient $A_\omega$ remains smaller than $e^{- 3/2}$.}
\label{coefs}
\end{figure}

In Fig.~\ref{coefs}, we represent the 4 coefficients of \eq{eq:Bsub} obtained by numerically solving \eq{eq:om}. On the horizontal axis, we use $\ln (\om/\ommax)$, where $\ommax$ is the theoretical maximal possible frequency for a counterpropagating wave sent by the wave maker. For the present series, it ranges from $2.17$ to $2.85 {\rm Hz}$.~\footnote{Notice that \eq{eq:waveeq} and the nonlinear equations of App. A of~\cite{Michel:2014zsa}, are left invariant under the rescaling $\om \to \lambda \om$, $x \to \lambda^{-2} x$, $h \to \lambda^{-2} h$, and $J \to \lambda^{-3} J$, which leaves $\Fmax$ unchanged. This invariance is lost when taking into account capillary effects~\cite{2013LNP...870...81R}.} On the vertical axis we show the logarithms of the norms of the four coefficients. From the two upper plots, giving respectively $\ln |\alpha_\om|$ (left) and $\ln |\beta_\om|$ (right), it is clear that the coefficients belong to two distinct classes depending on the trans- or sub-critical character of the flow. 

When the flow is sufficiently transcritical, the norms of $\alpha_\om$ and $\beta_\om$ grow as $(T_{\rm eff}/\om)^{1/2}$ for $\om \to 0$. For the flow with the highest value of $\Fmax \sim 1.25$, the relative difference between $T_{\rm eff}$ and the Hawking temperature $\kappa/2\pi$ (calculated at the white hole horizon on the downstream side) is less than $13\%$. Further simulations show that Hawking's predictions are recovered for a large range of frequencies  when $\Fmax \gtrsim 1.1$. (For localized obstacles, the finite size of the transcritical region leads to a suppression of $|\beta_\om|$ at ultra low frequencies not represented here, see~\cite{Michel:2014zsa}.) These results are in agreement with those of Refs.~\cite{Finazzi:2012iu,Robertson:2012ku} which were obtained by considering transcritical flows with $F(x)$ monotonically varying with $x$. The agreement can be understood on the basis that the emission spectrum is essentially fixed by the gradient on the downstream side.

Instead, for the three subcritical flows, $|\beta_\om|$ and $|\alpha_\om|$ vanish as $\om^{1/2}$ for $\om \to 0$~\cite{Michel:2014zsa}. In addition, one notices a sudden decrease of $|\alpha_\om|$ for $\om$ smaller than a critical frequency, that we call $\ommin$. It is given by the double root of the dispersion relation evaluated on top of the obstacle. For near critical flows $1- \Fmax \ll 1$, one finds 
\be
\ommin \simeq \frac{\cmin}{3 \hmin} (1 - \Fmax^2 )^{3/2}, 
\label{ommin}
\ee
where $\cmin = (g \hmin)^{1/2}$. As can be seen in the left lower panel, the norm of the transmission coefficient $|\tilde A_\om|$ is close to $1$ for $\om < \ommin$, here indicated by three vertical lines. For $\om < \ommin$, the waves are thus essentially transmitted. This is easily explained by considering the characteristics of \eq{eq:om}: the classical trajectories followed by counter propagating wave packets have no turning point for $\om < \ommin$. The large increase of transmission for $\om < \ommin$ was clearly verified in a dedicated experiment~\cite{Euve:2014aga}. On the right lower panel, for all flows, one observes that the norm of the ``gray body coefficient'' $A_\om$ remains $\lesssim e^{-3/2}$. Hence it plays no significant role in the scattering. 

In brief, we have shown that the scattering coefficients behave very differently in trans- and sub-sonic flows. For trans-critical flows with $\Fmax \gtrsim 1.1$, the situation is clear as we recover Hawking's prediction. As a consequence, since the temperature is fixed by the surface gravity $\kappa$, the upstream slope and the size of the trans-critical region on top of the obstacle play no significant role. For the other flows and in particular for sub-critical ones, to our knowledge, the relevant parameters which determine the emission spectrum have not been identified.

\subsection{Relevant parameters in sub-critical flows} 

\begin{figure}[h]
\begin{center}
\includegraphics[scale=0.625]{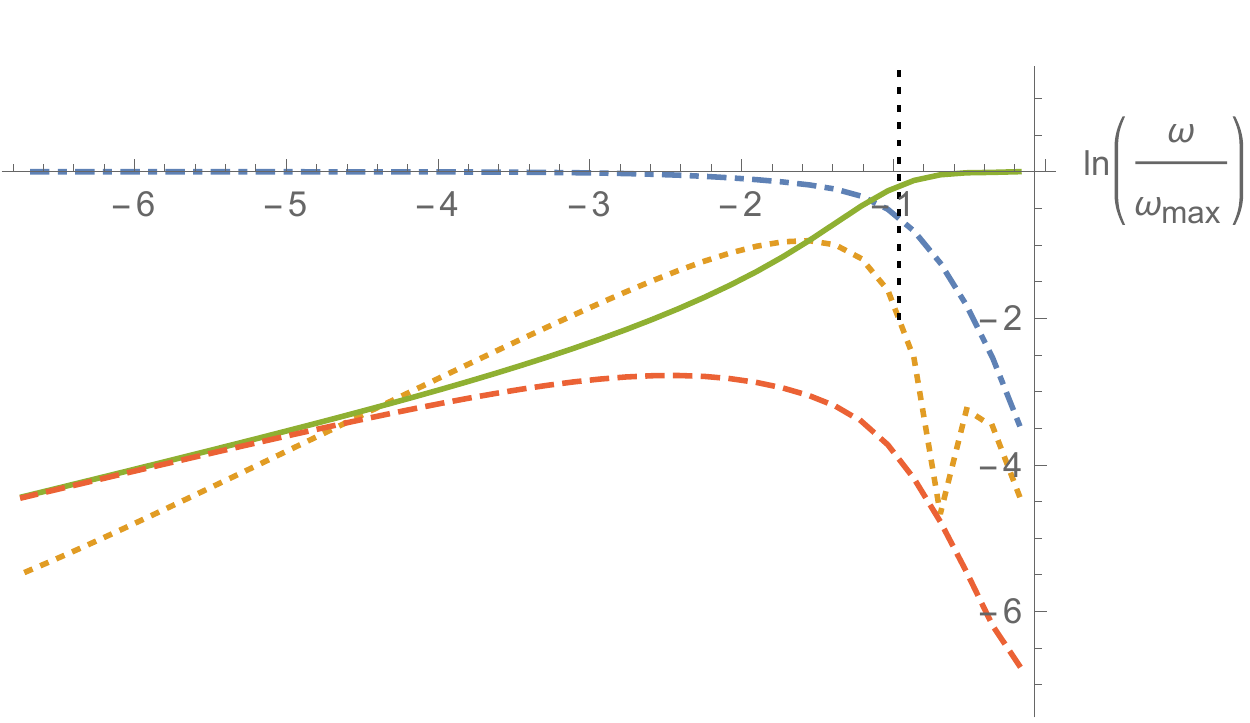} 
\includegraphics[scale=0.55]{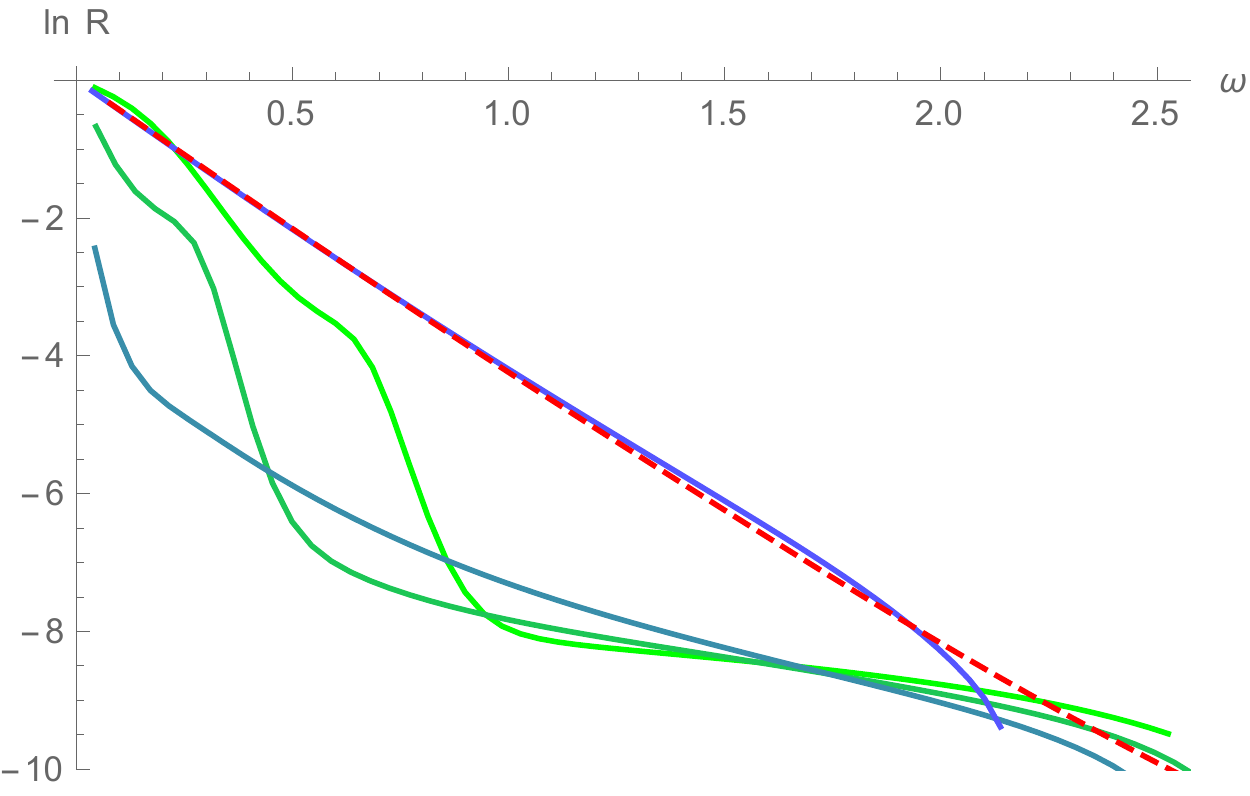}
\end{center}
\caption{On the left panel, we show the logarithm of the four scattering coefficients for the flow of the Vancouver experiment as functions of $\ln (\om/\ommax)$. The continuous, dashed, dashed-dotted, and dotted curves respectively correspond to $|\alpha_\om|$, $|\beta_\om|$, $|\tilde A_\om|$, and $|A_\om|$, see \eq{eq:Bsub}. The decrease of $|\alpha_\om|$ and the corresponding increase of the transmission coefficient $|\tilde A_\om|$ are both clearly observed for $\om < \ommin$, indicated by a vertical dotted line. On the right panel, we represent the logarithm of the ratio $R = |\beta_\om /\alpha_\om|^2$ for the Vancouver flow (in red dashed), for the three sub-critical flows, and that with highest $\Fmax$ of Fig.~\ref{coefs}. The last curve is almost straight, as expected since $|\beta_\om|^2$ follows a Planck spectrum, whereas the curves for the three sub-critical flows show oscillations. } 
\label{vanc}
\end{figure}

To guide our search, we present on the left panel of Fig.~\ref{vanc} the logarithms of the four scattering coefficients numerically evaluated for a flow similar to that used in the Vancouver experiment~\cite{Weinfurtner:2010nu}. The characteristic frequencies are $\ommax \sim 4.7 {\rm Hz}$, and $\ommin \sim 1.8 {\rm Hz}$. By comparison with the plots of Fig.~\ref{coefs}, one first sees that the behavior of these coefficients belongs to the class of sub-critical flows. In particular, the decrease of $|\alpha_\om|$ and the associated increase of $|\tilde{A}_\om|$ for $\om < \ommin$ are both clearly visible. In addition, we also see that $|\alpha_\om|^2 \sim |\beta_\om|^2 \propto \om$ for $\om \to 0$, which is less visible from the two upper plots in Fig.~\ref{coefs} but which is common to all our numerical simulations when considering subcritical flows. 

Yet, we observe some interesting differences. First, the higher value of $|\beta_\om|^2$ (with respect to that of the longer obstacle of Fig.~\ref{fig2} with the same value of $\Fmax$ and a similar slope on the downstream side) indicates that the higher slope of the Vancouver obstacle on its upstream side should play a significant role. Moreover, its narrower character should also explain why the decrease of the $|\alpha_\om|$ is less pronounced than that of the longer obstacle. We verified these two conjectures by varying the slope of the upstream side and the length of the plateau on top of the obstacle, while keeping the value of $\Fmax$ and the downstream slope fixed, see Appendix~\ref{Apa}. In Appendix~\ref{Apb} we show profiles of perturbations of the free surface to illustrate the difficulties to extract the transmission coefficient at low frequencies. 

It is also interesting to study the behavior of the quantity that was observed in the experiment of~\cite{Weinfurtner:2010nu}, namely the logarithm of $R_\om = |\beta_\om /\alpha_\om|^2$ associated with the two dispersive waves, see \eq{eq:Bsub}. On the right panel of Fig.~\ref{vanc}, as a function of $\om$ (in Hz), we represent $\ln R_\om $ (numerically computed using \eq{eq:om}) for the Vancouver flow (in dashed), and for four flows of Fig.~\ref{coefs}, namely the transcritical flow with highest $\Fmax$ and the three sub-critical flows. As noticed in~\cite{Michel:2014zsa}, $\ln R_\om$ numerically computed for the Vancouver flow is almost linear, in agreement with what was observed in~\cite{Weinfurtner:2010nu}. This is surprising since $|\beta_\om|^2$ is highly suppressed, see the dashed curve on the left panel. In fact, $\ln R_\om$ closely follows the one evaluated for the transcritical flow, which is there linear because $|\beta_\om|^2$ is Planckian, and $|\alpha_\om|^2 - |\beta_\om|^2 = 1$ is well satisfied. The important lesson is that the behavior of $\ln R_\om$ is unable to differentiate between the radically distinct behaviors of $|\alpha_\om|^2$ and $|\beta_\om|^2$ for these two flows. The same lesson applies to flows where $F$ monotonically varies with $x$, see Fig.~6 of~\cite{Michel:2014zsa}.

\newpage

Yet, it would be nice to understand why $\ln R_\om$ for the Vancouver flow is almost linear in $\om$.~\footnote{We are grateful to Bill Unruh for pointing out to us the importance of this question.} On the right panel of Fig.~\ref{vanc}, we notice that the sub-critical flow with the lowest value of $\Fmax\simeq 0.75$, comparable to that of the Vancouver flow, has a {\it larger} value of $\ln R_\om$ when compared to the other two subcritical flows, even though its emission spectrum $|\beta_\om|^2$ is smaller, see right upper panel in Fig.~\ref{coefs}. This reveals that, when lowering the value of $\Fmax$, $|\alpha_\om|^2$ decreases faster than $|\beta_\om|^2$ so as to increase the slope of $\ln R_\om$. We then notice that for very low frequency the slope of $\ln R_\om$ for the sub-critical flow agrees with that of the Vancouver flow before decreasing and showing oscillations. This suggests that a more linear behavior would be obtained if narrowing the obstacle, as the oscillations result from interferences between the scattering from its two slopes, which are here well separated. (In fact, such oscillations in the spectrum were already observed in~\cite{Finazzi:2010yq} when considering flows with two scattering zones, see also the black hole laser for similar interfering effects~\cite{Coutant:2009cu}.) In Appendix~\ref{Apa}, we verify that this is the case. In brief, we conjecture that the linear slope of $\ln R_\om$ observed in~\cite{Weinfurtner:2010nu} is due to the rather low value of $\Fmax \sim 0.75$ and the relatively narrow character of the obstacle, which is small enough to avoid interferences between scattering on the upstream slope and the downstream one. 

\section{Conclusions}

By studying the variations of the scattering coefficients along series of background flows, we made the following important observations. Firstly, we showed that the scattering coefficients principally depend on $\Fmax$, the maximal value of the Froude number reached on top of the obstacle. Secondly, Hawking's prediction are only recovered for transcritical flows with $\Fmax \gtrsim 1.1$. Thirdly, for subcritical flows, we showed that varying separately the length and each slope of the obstacle significantly affect the scattering coefficients. Fourthly, we observed that oscillations of $\ln R_\om$ found for long obstacles are suppressed for narrower ones. As a final comment, we wish to point out that although derived in the context of water flows, these results should also apply (up to minor modifications) to other systems such as flowing atomic condensates~\cite{Garay:1999sk}, polariton systems~\cite{Gerace:2012an}, and non-linear optical media~\cite{Philbin:2007ji}. 

\medskip
\medskip

{\bf Acknowledgements.} The authors are grateful to L-P. Euv\'e and G. Rousseaux for many discussions. They are also grateful to the participants of the 1-day QEAGE workshop help in LPT-Orsay on July $1^{st}$ for interesting comments. They thank the organizers and the participants of the parallel session  ``Regular and Analogue Black Holes'' of the M. Grossmann meeting held in Rome in July 2015. Finally they thank W.~Unruh, S.~Weinfurtner and S.~Robertson for useful comments on an earlier version of this manuscript. This work received support from the French National Research Agency under the Program Investing in the Future Grant No. ANR-11-IDEX-0003-02 associated with the project QEAGE (Quantum Effects in Analogue Gravity Experiments). 

\newpage

\begin{appendices} 

\section{\writeadot Length of the obstacle and linearity of ${\rm ln} R$} 
\label{Apa}

\begin{figure}[h]
\begin{center} 
\includegraphics[width=0.49 \linewidth]{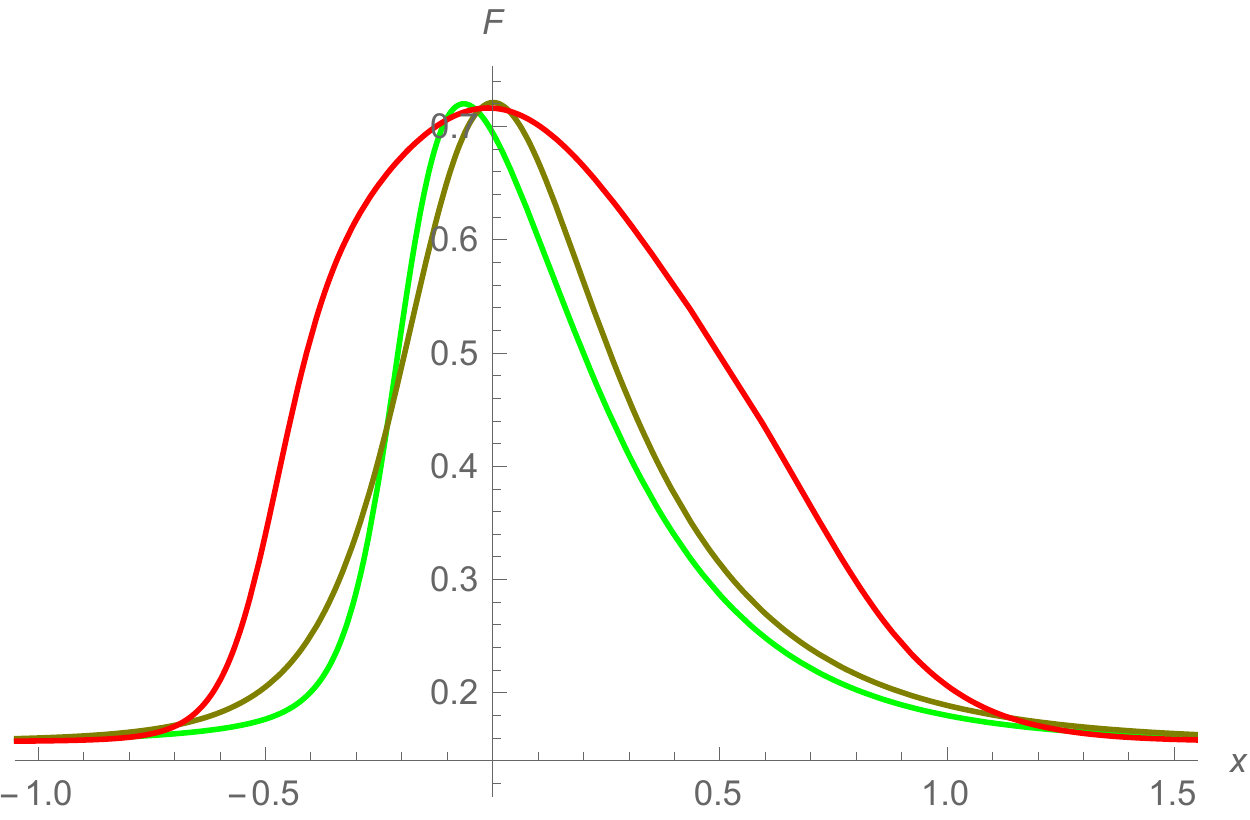}
\includegraphics[width=0.49 \linewidth]{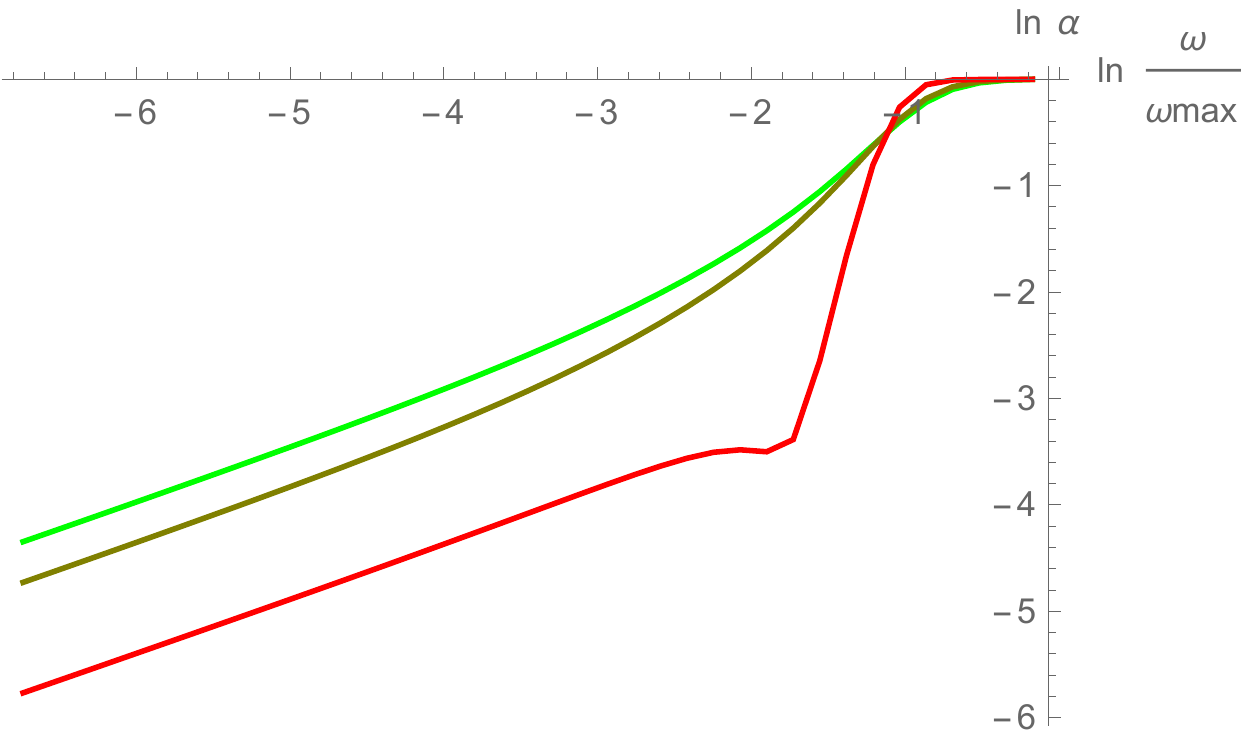}
\includegraphics[width=0.49 \linewidth]{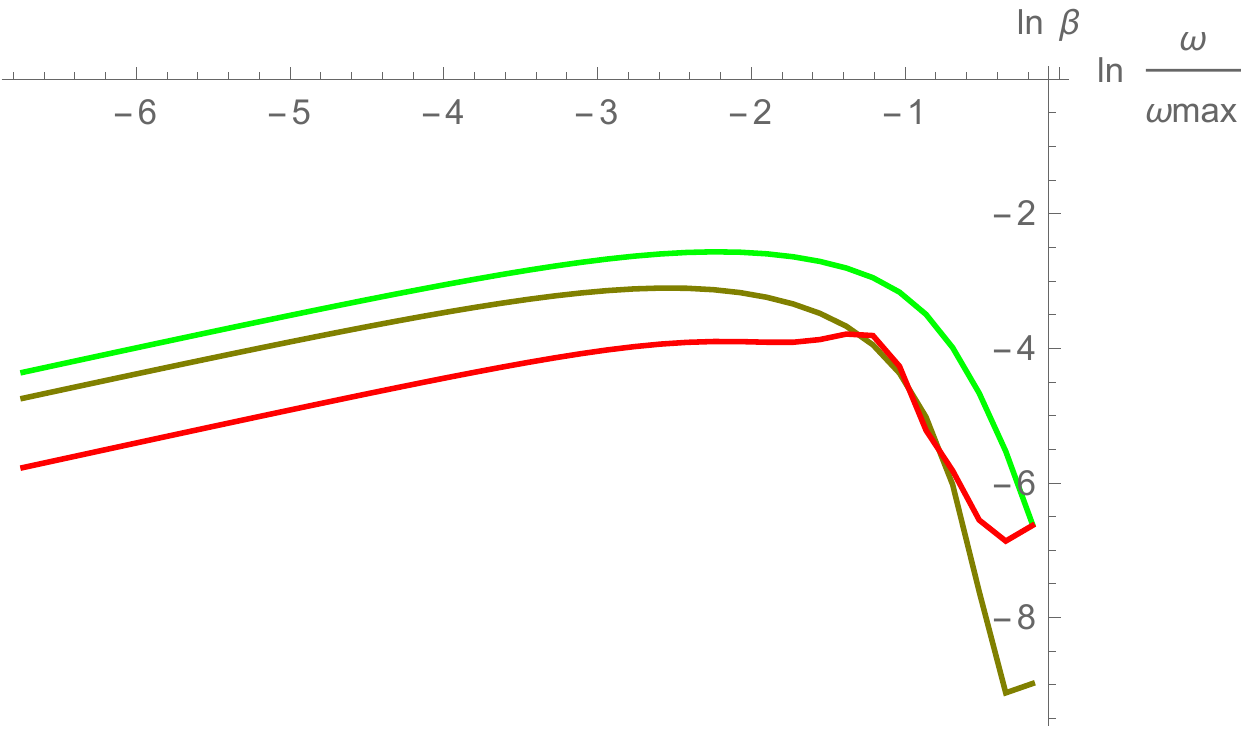}
\includegraphics[width=0.49 \linewidth]{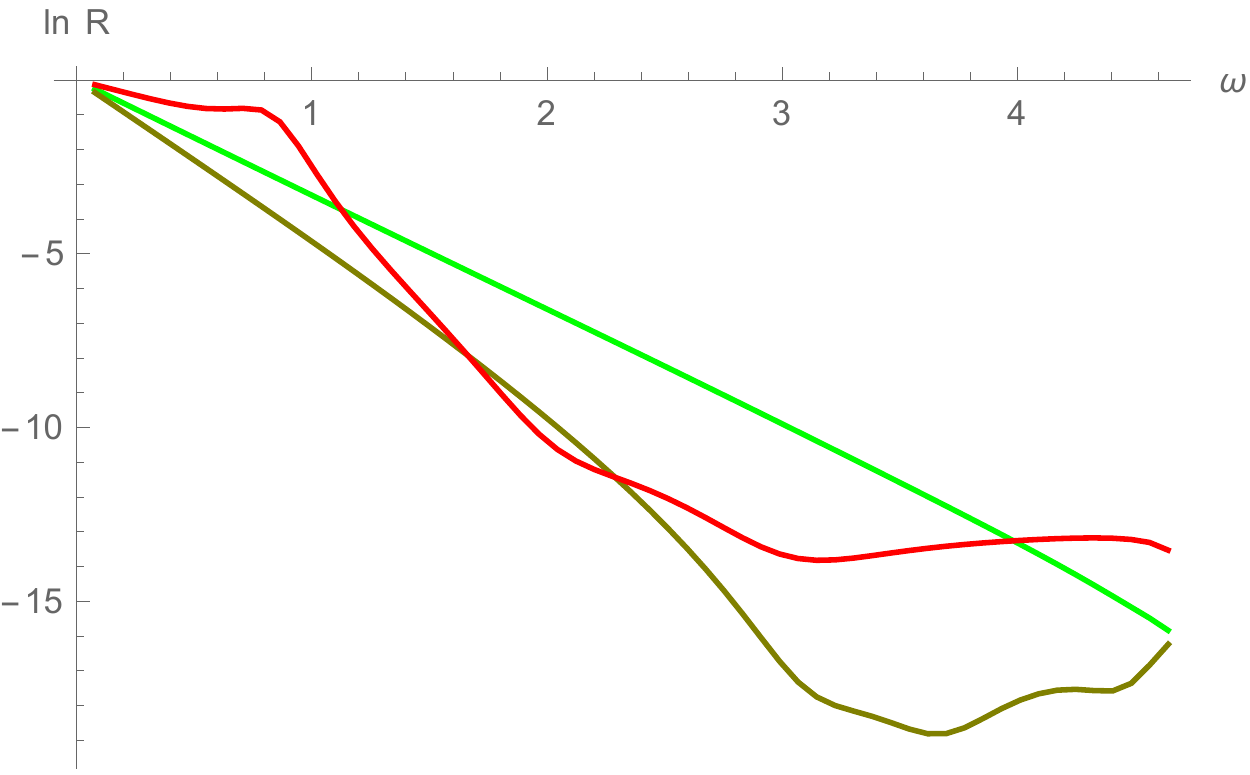} 
\end{center} 
\caption{For three flows with the same value of $\Fmax$, and approximatively the same 
downstream (right) slope, we show the Froude number (top, left) as a function of $x$, the scattering coefficients $\ln |\alpha_\om|$ (top, right) and $\ln | \beta_\om|$ (bottom, left) as functions of $\ln (\om / \ommax)$, and $\ln R_\om$ (bottom, right) as a function of $\om$. The green curves are obtained with an obstacle similar to the one used in~\cite{Weinfurtner:2010nu}. The brown curves correspond to an obstacle with a smaller ascending slope, 
and the red ones to a longer obstacle. From the lower plots we clearly see that reducing the length of the obstacle, and increasing the downstream slope, increase the power $|\beta_\om|^2$ and the linearity of ${\rm ln} \, R_\om$, even though one gets $|\beta_\om|^2 \propto \om$ as $\om \to 0$ for the three cases. 
} \label{slope}
\end{figure}

We study the behavior of the scattering coefficients in subcritical flows when widening the obstacle by increasing the length of the plateau on its top and by decreasing its upstream slope. In the upper left plot of \Fig{slope}, we show the profiles of $F(x)$ of the three flows. We see that their value of $\Fmax \simeq 0.72$ is unchanged, and that their downstream (right) slope is essentially the same. The narrowest obstacle (dark green) has a length and slopes close to those of the obstacle used in the Vancouver experiment. 

In the upper right plot, we observe that the slope of $|\alpha_\om|$ for $\om < \ommin$ significantly increases when increasing the size of the obstacle. This is expected since the transition between wave-blocking ($\om > \ommin$) and transmission is smoother for narrower obstacles. In the lower left plot, the 3 curves representing $\ln |\beta_\om|$ establish that, although $|\beta_\om|^2 \propto \om$ as $\om \to 0$ for the three flows, the reduction of the length of the obstacle, and the increase of the upstream slope, both increase the mode mixing for small frequencies.  From this we conclude that, for subcritical flows, the downstream slope (of the ``white hole'' horizon) alone cannot explain the emission spectrum. This is confirmed by examining the behavior of the quantity observed in~\cite{Weinfurtner:2010nu}, see the lower right plot where $\ln R_\om$ is shown as a function of $\om$. We notice that reducing the length of the upper part of the obstacle significantly increases the linearity of $\ln R$ in $\om$. This can be understood from the fact that the radius of curvature for the narrowest obstacle is of the same order as the dispersive length $\hmin$. We conjecture that the absence of a clear scale separation between the values of the inverse gradient of the flow and the dispersive length $\hmin$ for the Vancouver obstacle explains the linearity of $\ln R_\om$. 

We have performed extra simulations (not represented) where we multiplied, and divided, by 2 each slope. We found that decreasing the upstream slope monotonically and significantly decreases the slope of $\ln R_\om$. Instead, decreasing the downstream slope slightly {\it increases} that of $\ln R_\om$, while increasing it produces oscillations in  $\ln R_\om$, probably because interferences between the two sides are enhanced. These simulations confirm that the upstream slope plays a leading role for this type of obstacles. It would be important to see if these properties (obtained from \eq{eq:om}) are validated in forthcoming experiments.

\section{\writeadot Profiles of perturbations of the free surface}
\label{Apb}

\begin{figure}[h] 
\begin{center}
\includegraphics[width = 0.49 \linewidth]{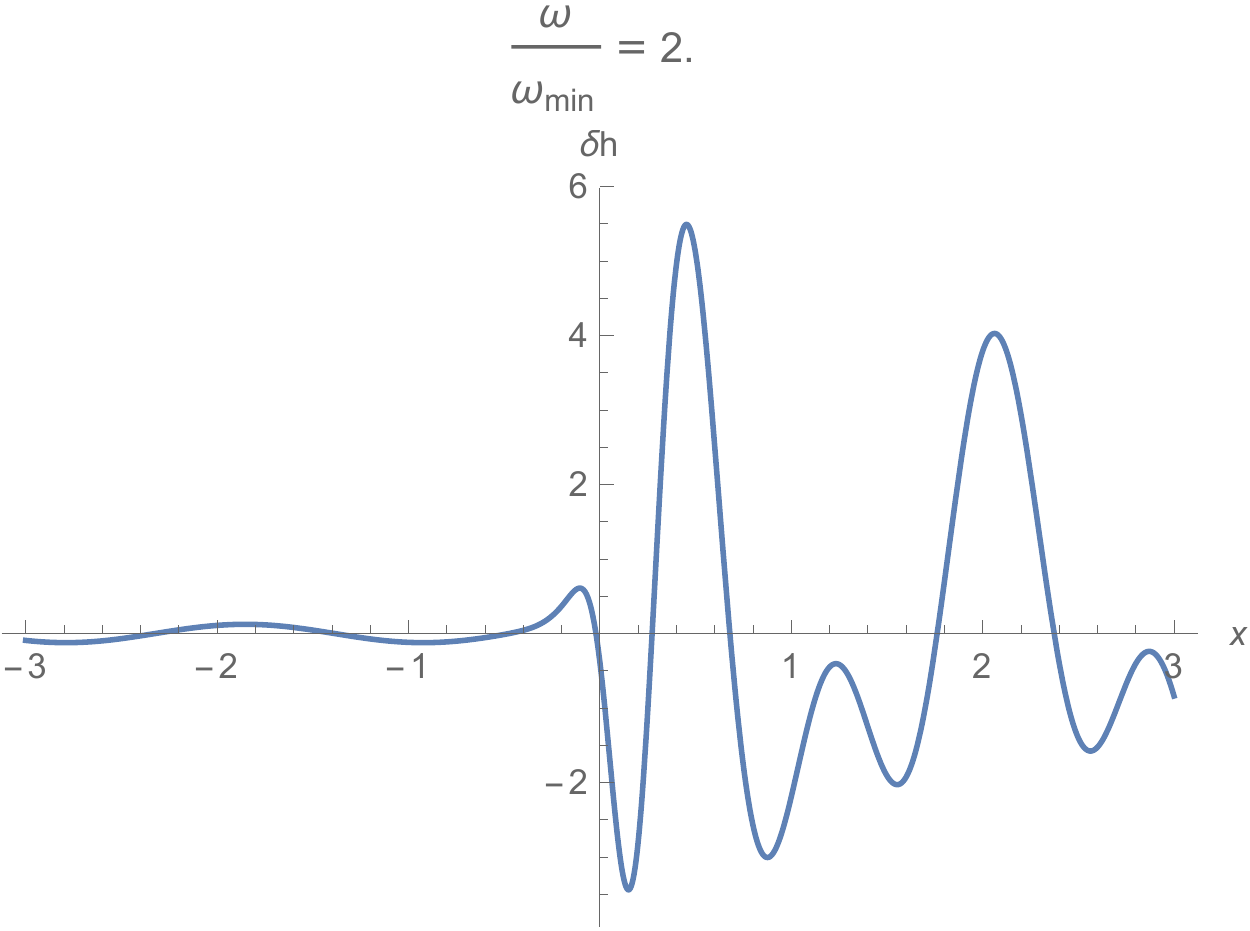} 
\includegraphics[width = 0.49 \linewidth]{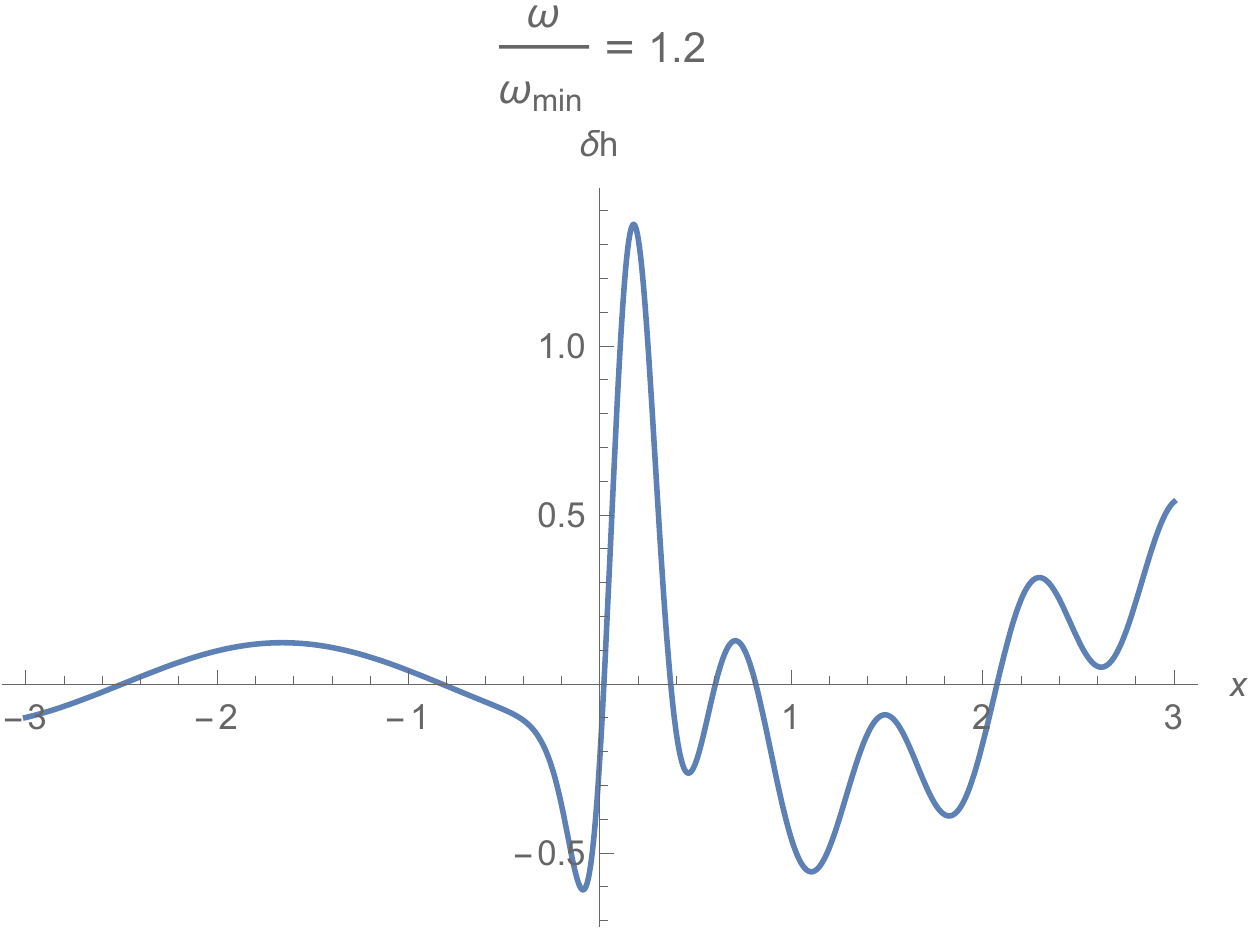} 
\includegraphics[width = 0.49 \linewidth]{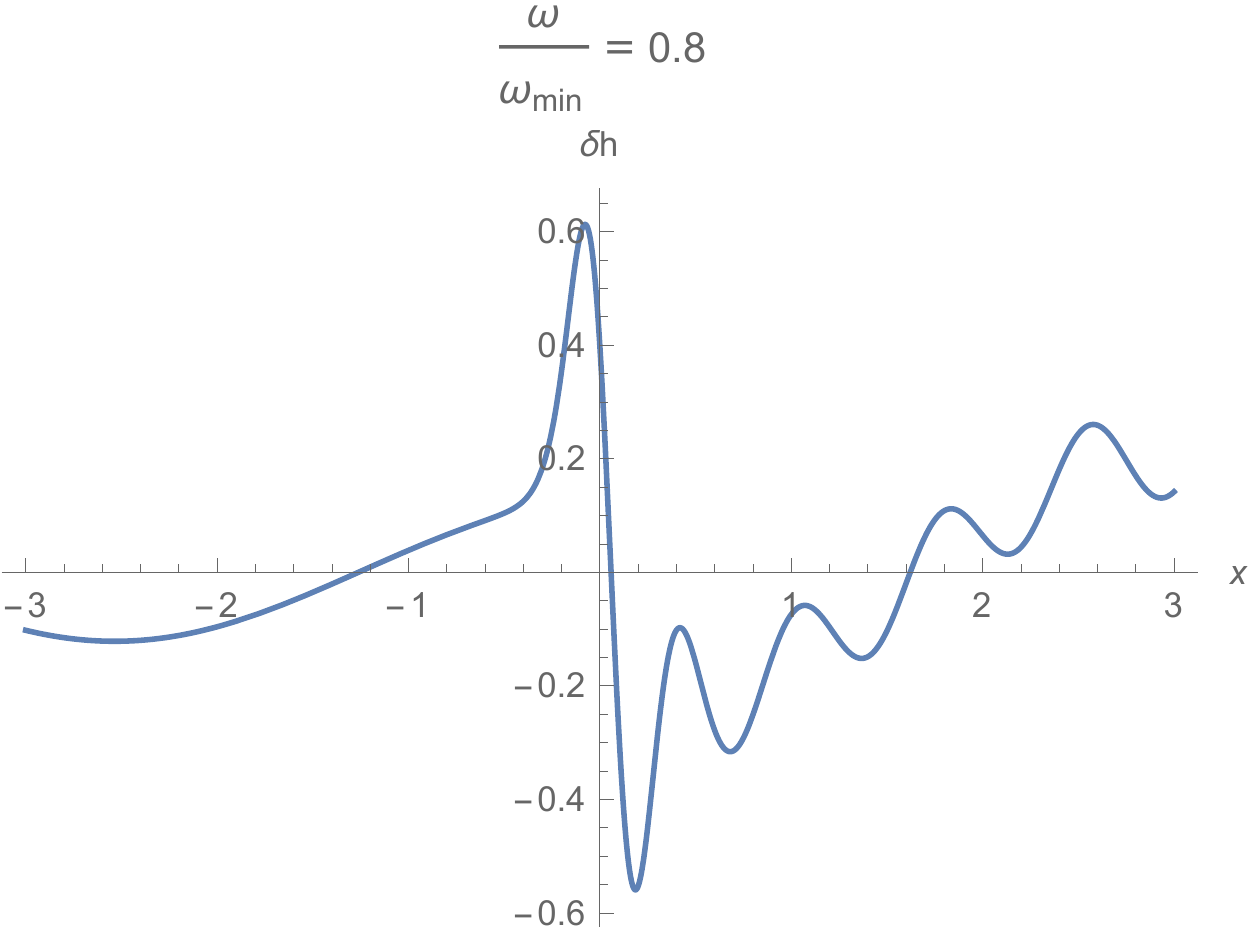}
\includegraphics[width = 0.49 \linewidth]{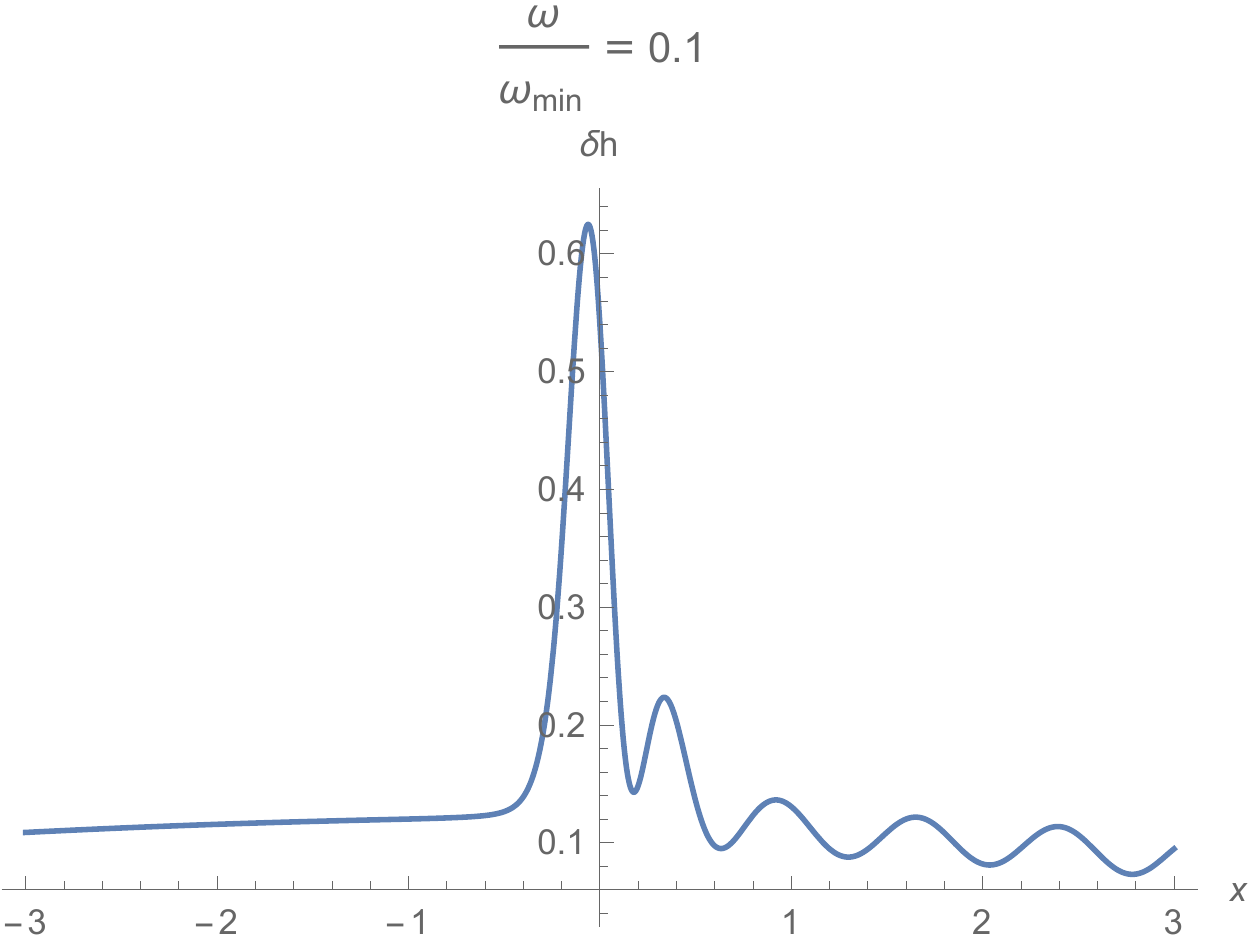}
\end{center} 
\caption{Profile of the free-surface variation $\delta h_\om$ for the incoming counter-propagating mode scattered on a flow similar to that used in Vancouver, and evaluated at four frequencies $\om = 2 \ommin$ (top, left), $\om = 1.2 \ommin$ (top, right), $\om = 0.8 \ommin$ (bottom, left), and $\om = 0.1 \ommin$ (bottom, right). The corresponding values of $|\tilde{A}_\om|$ are respectively $0.055$, $0.36$, $0.70$, and $0.99$. The mode is normalized 
by imposing that the amplitude of the transmitted mode is independent of $\om$, which means that the incoming wave amplitude remains finite in the limit $\om \to 0$. } \label{profiles}
\end{figure}

We illustrate the difficulty of extracting the transmission coefficient $\tilde A_\om$ from measurements of the free surface at low frequency, which means very long wave lengths. In \Fig{profiles} we present profiles of $\delta h_\om(x)$, the linear variation of the free surface for the incoming counterpropagating mode of \eq{eq:Bsub} evaluated at 4 different frequencies. To see the transmitted amplitude, $\delta h_\om(x)$ is evaluated at a time when its amplitude in the downstream region reaches its maximum value.
 We used the parameters of~\cite{Weinfurtner:2010nu}, for which $\ommin \approx 1.8$ Hz. On the right incoming side, the profile is complicated as 4 waves interfere. On the left it is simple as there is only the transmitted wave. When $\om$ is significantly larger than $\ommin$, one easily sees that the transmitted mode has a very small amplitude. When decreasing $\om$ close to or below $\ommin$, this amplitude grows and becomes comparable to that of the dispersive waves. However, as the corresponding wave-length becomes very large (it is already of a few meters for $\om = \ommin$, and goes to infinity in the limit $\om \to 0$), it is not easily seen. In fact, on the lower right plot, we notice that the incoming mode raises the height of the free surface on the right side by about the same amount as the transmitted amplitude. 

On the lower right plot, we also see that the typical amplitude of the short wave length modulations of $\delta h_\om$ on the right side is of the {\it same order} as the transmitted amplitude on the left, as if reflection was still significant. However, for $\om = 0.1 \ommin$ one has $|\tilde{A}_\om| \simeq 0.99$ %and 
while %F
the coefficients $|\alpha_\om|$ and  $|\beta_\om|$ in front of the dispersive waves are both {\it much smaller} than $1$, as can be seen on the left panel of Fig.~\ref{vanc}. The reason for this discrepancy is as follows. The kinematical relations between $\delta h^{\rm disp}_\om$ and $|\alpha_\om|$ ($|\beta_\om|$) on one side, and between $\delta h^{\rm transm}_\om$ and $|\tilde{A}_\om|$ on the other side, imply that
\be
\frac{\delta h^{\rm disp}_\om}{\delta h^{\rm transm}_\om}  \propto  \frac{\alpha_\om}{\om^{1/2}\, \tilde{A}_\om},
\ee
see the discussion after Eq.~(30) in~\cite{Michel:2014zsa}. Hence, when the ratio $\delta h^{\rm disp}_\om/\delta h^{\rm transm}_\om$ goes to constant for $\om \to 0$, as predicted by our numerical simulations, $|\alpha_\om |$ and $|\beta_\om|$ both vanish as $\om^{1/2}$. These features might explain why transmission below $\ommin$ was not reported in~\cite{Weinfurtner:2010nu}.

\end{appendices} 

\bibliographystyle{myunsrt}
\bibliography{bibliopubli}

\begin{thebibliography}{10}

\bibitem{Unruh:1980cg}
W.G. Unruh.
\newblock {\em Phys. Rev. Lett.}, 46:1351--1353, 1981.

\bibitem{Unruh:1994je}
W.G. Unruh.
\newblock {\em Phys. Rev. D}, 51:2827--2838, 1995.

\bibitem{Schutzhold:2002rf}
R.~Schutzhold and W.~G. Unruh.
\newblock {\em Phys. Rev. D}, 66:044019, 2002.

\bibitem{Brout:1995wp}
R.~Brout, S.~Massar, R.~Parentani, and P.~Spindel.
\newblock {\em Phys. Rev. D}, 52:4559--4568, 1995.

\bibitem{Corley:1996ar}
S.~Corley and T.~Jacobson.
\newblock {\em Phys. Rev. D}, 54:1568--1586, 1996.

\bibitem{Macher:2009tw}
J.~Macher and R.~Parentani.
\newblock {\em Phys. Rev. D}, 79:124008, 2009.

\bibitem{Finazzi:2012iu}
S.~Finazzi and R.~Parentani.
\newblock {\em Phys. Rev. D}, 85:124027, 2012.

\bibitem{Coutant:2011in}
A.~Coutant, R.~Parentani, and S.~Finazzi.
\newblock {\em Phys. Rev. D}, 85:024021, 2012.

\bibitem{Robertson:2012ku}
S.~J. Robertson.
\newblock {\em J. Phys. B}, 45:163001, 2012.

\bibitem{Rousseaux:2007is}
G.~Rousseaux, C.~Mathis, P.~Maissa, T.~G. Philbin, and U.~Leonhardt.
\newblock {\em New J. Phys.}, 10:053015, 2008.

\bibitem{Weinfurtner:2010nu}
S.~Weinfurtner, E.~W. Tedford, M.~C.J. Penrice, W.~G. Unruh, and G.~A.
  Lawrence.
\newblock {\em Phys. Rev. Lett.}, 106:021302, 2011.

\bibitem{Belgiorno:2010wn}
F.~Belgiorno, S.L. Cacciatori, M.~Clerici, V.~Gorini, G.~Ortenzi, et~al.
\newblock {\em Phys. Rev. Lett.}, 105:203901, 2010.

\bibitem{Schutzhold:2010am}
R.~Schutzhold and W.~G. Unruh.
\newblock {\em Phys. Rev. Lett.}, 107:149401, 2011.

\bibitem{Michel:2014zsa}
F.~Michel and R.~Parentani.
\newblock {\em Phys.Rev.}, D90(4):044033, 2014.

\bibitem{Busch:2014hla}
X.~Busch, F.~Michel, and R.~Parentani.
\newblock {\em Phys.Rev.}, D90(10):105005, 2014.

\bibitem{Unruh:2012ve}
W.~G. Unruh.
\newblock {\em Lect. Notes Phys.}, 870:63--80, 2013.

\bibitem{Coutant:2012mf}
A.~Coutant and R.~Parentani.
\newblock {\em Phys.Fluids}, 26:044106, 2014.

\bibitem{Euve:2014aga}
L.-P. Euvé, F.~Michel, R.~Parentani, and G.~Rousseaux.
\newblock {\em Phys. Rev.}, D91(2):024020, 2015.

\bibitem{OnOverReflection}
D.~J. {Acheson}.
\newblock {\em Journal of Fluid Mechanics}, 77:433--472, October 1976.

\bibitem{2013LNP...870...81R}
G.~{Rousseaux}.
\newblock {The Basics of Water Waves Theory for Analogue Gravity}.
\newblock In D.~{Faccio}, F.~{Belgiorno}, S.~{Cacciatori}, V.~{Gorini},
  S.~{Liberati}, and U.~{Moschella}, editors, {\em Lecture Notes in Physics,
  Berlin Springer Verlag}, volume 870 of {\em Lecture Notes in Physics, Berlin
  Springer Verlag}, page~81, 2013.

\bibitem{Finazzi:2010yq}
S.~Finazzi and R.~Parentani.
\newblock {\em Phys. Rev. D}, 83:084010, 2011.

\bibitem{Coutant:2009cu}
A.~Coutant and R.~Parentani.
\newblock {\em Phys. Rev. D}, 81:084042, 2010.

\bibitem{Garay:1999sk}
L.~J. Garay, J.~R. Anglin, J.~I. Cirac, and P.~Zoller.
\newblock {\em Phys. Rev. Lett.}, 85:4643--4647, 2000.

\bibitem{Gerace:2012an}
D.~Gerace and I.~Carusotto.
\newblock {\em Phys. Rev. B}, 86:144505, 2012.

\bibitem{Philbin:2007ji}
T.~G. Philbin, C.~Kuklewicz, S.~Robertson, S.~Hill, F.~Konig, and U.~Leonhardt.
\newblock {\em Science}, 319:1367--1370, 2008.

\end{thebibliography}

\end{document}